\pdfoutput=1

\documentclass[10pt,conference]{IEEEtran}

\usepackage{amsmath,epsfig}
\usepackage{amssymb}
\usepackage{amsfonts}
\usepackage{array,dcolumn}
\usepackage{psfrag}
\usepackage{graphicx}
\usepackage{psfrag}
\usepackage{amsmath}
\usepackage{amsfonts}
\usepackage{amssymb}
\usepackage{accents}
\usepackage{cite}
\usepackage{subfigure}
\usepackage{pstricks,exscale}
\usepackage{dsfont}
\usepackage{float}
\usepackage{verbatim}

\begin{document}

\title{Energy-efficiency in Decentralized Wireless Networks: A Game-theoretic Approach inspired by Evolutionary
Biology}

\author{Andrej Gajduk $^*$, Zoran Utkovski$^{\#,*}$, Lasko Basnarkov $^{*,\dagger}$ and Ljupco Kocarev $^{*,\dagger,\ddagger}$\\
$^*$ Laboratory for Complex Systems and Networks,\\
Macedonian Academy of Sciences and Arts, Republic of Macedonia\\
${^\#}$Faculty of Computer Science, University Goce Delcev, Stip,
Republic of Macedonia%
\\
${^\dagger}$Faculty of Computer Science and Engineering, University SS. Cyril and Methodius Skopje, \\
Republic of Macedonia%
\\
${^\ddagger}$BioCircuits Institute, University of California San Diego, USA}%

\maketitle

\begin{abstract}
Energy efficiency is gaining importance in wireless communication
networks which have nodes with limited energy supply
and signal processing capabilities. 
We present a numerical study of cooperative communication scenarios
based on simple local rules. This is in contrast to most of the
approaches in the literature which enforce cooperation by using
complex algorithms and require strategic complexity of the network
nodes. The approach is motivated by recent results in evolutionary
biology which suggest that, if certain mechanism is at work,
cooperation can be favored by natural
selection, i.~e. even selfish actions 
of the individual nodes can lead to emergence of cooperative
behavior in the network. The results of the simulations in the
context of wireless communication networks verify these
observations and indicate that uncomplicated local rules, followed
by simple fitness evaluation, can generate network behavior which
yields global energy efficiency.

\end{abstract}

\section{Introduction}

One of the major aspects of future wireless networks is their
energy efficiency, which lies at the focus of this work. The
provisioning of energy efficient protocols and communication
schemes is one of the main challenges in the design of present and
future communication networks. The concept of energy efficiency is
particularly relevant in emerging heterogeneous networks which,
besides the "classical" communication nodes, include various other
devices with low-power capabilities, such as sensors and other
nodes producing machine-type traffic.

Wireless communication systems have two fundamental properties.
The first one is that they receive power decays according to a
power law function of the distance between the users, which puts
stress on the power consumption. The second important property is the broadcast
nature of the wireless communication, which leads to interference
between the users. With the increase in the number of subscribers
and growth in data traffic in wireless networks, these two
features gain in importance and have a strong adverse effect on
the network performance in terms of throughput and energy
consumption.

The study of the fundamental limits of wireless networks suggests
that cooperation among the units could potentially overcome these
effects. In this context, techniques such as cooperative diversity
\cite{sendonaris2003user1, sendonaris2003user2} and interference
alignment \cite{cadambe2008} have been proposed.

As cooperation is considered to be beneficial in general, most of
the present approaches which deal with the aspects of cooperative
communications, assume that the network nodes act in a pre-defined
way, i.~e. their behavior is determined by (usually) centralized
network rules. In such systems it is often assumed that the
cooperation between the network nodes is beneficial "by default"
and no freedom is left to the individual nodes to decide about
their involvement in the cooperative act. Besides the networks
with central infrastructure, this is also the case in
decentralized networks such as ad-hoc networks. One general
observation is that the proper functioning of these networks is
generally maintained either by enforcing cooperation, or by
keeping track of the cooperative behavior which demands intensive
computation
\cite{zhong2003sprite,buchegger2003wiopt,liu2003reputation,
anantvalee2007reputation,mundinger2008analysis}. Since cooperation
is associated with a cost (usually energy) and requires certain
signal processing capabilities (computational complexity), this
approach may lead to a "cooperation burden" which can be
unreasonably high for some network nodes.

The above described networks, as well as all other present
communication networks, can be seen as systems which are mostly
``engineered'', meaning that they reflect a number of accepted
principles of good design. The parts in these systems have known
functions and designers attempt to maintain separation of
concerns. On the other hand, there are other networks such as
biological, social and economic, which evolve over time as a
result of the interactions between the system entities and the
environment. A formidable remark is that, compared to present
communication systems, information storage and processing in
living organisms is more efficient by many orders of magnitude,
with respect to both information density and energy consumption.
Low energy consumption paired with simplicity, efficiency and
adaptability seems to be an important objective for information
exchange in living organisms. 

Motivated by these observations, we will present a game-theoretic
approach to the energy efficiency in decentralized wireless
networks, which is motivated by the insights obtained from
evolutionary biology. In particular, we will concentrate on the
concept of cooperation and its emergence in communication networks. 
The model studied in this work considers forwarding the packets of
the senders towards the receivers by entities that are
cooperators. The choice between cooperative or defective behavior
that nodes make in this game theoretic approach is based solely on
calculation of their energy consumption. This assumption greatly reduces
 the computation complexity, as compared with the studies
found in the literature \cite{buchegger2005self,liu2003reputation,
anantvalee2007reputation,mundinger2008analysis}. The simulations
show that the population of cooperators persists although the
nodes' decisions are selfish, which results in decreased average
energy consumption both for the whole population and for the
individual nodes.

The remaining of the paper is organized as follows. In
Section~\ref{sec:energy_efficiency_and_cooperation} the relation
between energy efficiency and cooperation is given. Next, in
Section~\ref{sec:network_model} the model studied in this work is
described. The experiments and the results obtained are explained
in Section~\ref{sec:experiment_setup}.
Section~\ref{sec:conclusion} concludes this paper.

\section{Energy Efficiency and Cooperation}
\label{sec:energy_efficiency_and_cooperation}

\subsection{Wireless communication networks}
The primary focus of this paper is on the communication in
wireless networks. However, the key ideas could be applicable to other
communication networks, as well as networks in general, including
social and economic networks.


The performance analysis of wireless networks is often based on
simplifying assumptions. As a general rule, the cost of
establishing cooperation in wireless networks is not properly
taken into account when deriving the performance limits of
different cooperative schemes \cite{lozano2013fundamental}. For example, in
some scenarios the benefits of cooperation might be overshadowed
by the cost of establishing cooperation in the first place. Also,
very often a central infrastructure/control is assumed, which is
not always the case. In addition, as cooperation comes at a cost
for the network users, in a network which lacks centralized
control, at certain time instants for some users it might be
beneficial to defect, instead of cooperate.

Cooperation in decentralized networks is usually established by
complex algorithms
\cite{buchegger2002performance,michiardi2002core}, which usually
promote/enforce cooperation based on reputation tables about the
users' behavior. In contrast to the present approaches which rely
on complex algorithms in order to enforce cooperation, we are
interested in cooperation which emerges as a result of the system
evolution. This approach is inspired by recent results in
evolutionary biology which suggest that cooperation can emerge and
persist in evolving systems, i.~e. that cooperation can also be
favored by natural selection, if certain mechanism is at work
\cite {nowak2006, nowak2006five}.

While we will build on the legacies of communication protocols for
establishing cooperation in decentralized networks, our approach
differs in one important aspect. Namely, we will not assume
cooperation to be beneficial "by default", but we will rather
adopt a game-theoretic approach where the network nodes
\textit{decide} whether to cooperate or not based only on their
individual fitness, where the fitness is a quantity related to the energy
 consumption of the individual nodes. We will propose and
evaluate different strategies of the individual nodes in terms of
the average energy consumption. The focus will be on simple,
decentralized strategies which do not require strategic complexity
of the involved nodes. This is in the spirit of evolving systems
(such as biological) where the cooperative behavior is based on
simple rules and its emergence can be understood by relatively
simple mechanisms. As we will see, some interesting insights will
appear as consequence of the analysis performed in this paper. The
most important one is probably the conclusion that, under certain
circumstances, there are simple strategies which do not enforce
cooperation, but it emerges as a result of the network evolution.
These results also serve as an indicator
that uncomplicated local, evolutionary-like rules, followed by
simple fitness evaluation, can generate network behavior which
yields global energy efficiency.


\subsection{Cooperation in biological systems} Recent results in
biology \cite{axelrod1981evolution, nowak2006, nowak2006five,
cremer2012} show that cooperation has played a fundamental role in
many of the major transitions in biological evolution and is
essential to the functioning of a large number of biological
systems. Observations show that cooperative interactions are
required for many levels of biological organization ranging from
single cells to groups of animals. Human society, as well, is
based to a large extent on mechanisms that promote cooperation. In
the following paragraphs we will shortly address the concept of cooperation
in biology and revisit the candidate mechanisms which explain the
emergence and stability of cooperation.

\subsubsection{Emergence of cooperation in biological systems} It is
well known that in unstructured populations, natural selection
favors defectors over cooperators. There is much current interest,
however, in studying evolutionary games in structured populations
and on graphs \cite{nowak2006}. In \cite{nowak2006} the authors
describe a simple rule that is a good approximation for different
graphs, including cycles, spatial lattices, random regular graphs,
random graphs and scale-free networks. The conclusion is that
natural selection favors cooperation, if the benefit of the
altruistic act, $b$, divided by the cost, $c$, exceeds the average
number of neighbors, $k$, $b/c>k$. The intuition behind is that in
this case cooperation can evolve as a consequence of "social
viscosity" even in the absence of reputation effects or strategic
complexity.
\subsubsection{Mechanisms behind the emergence
of cooperation} Candidate mechanisms in biology which are able to
explain the emergence and stability of cooperation are kin
selection, direct reciprocity, indirect reciprocity, network
reciprocity, and group selection \cite{nowak2006five}.

Among the candidate mechanisms which promote cooperation based on
natural selection, we identify network reciprocity as the most
relevant for wireless communication networks.
Network reciprocity is a mechanism that aims to explain why cooperation persists in populations where some individuals interact more often~\cite{nowak2006five}.
The approach
of capturing this effect is evolutionary graph theory, which
allows the study of how spatial structure affects evolutionary
dynamics. According to this model, the individuals of a population
occupy the vertices of a graph, where the edges determine who
interacts with whom. Additionally, the users are assumed to be
plain cooperators and defectors without any strategic complexity.
In this setting, the experiments show that cooperators can prevail
by forming network clusters, where they help each other. 

\section{Energy-efficient Decentralized Wireless Networks}
\label{sec:network_model}

Many of the most fundamental instances of cooperation in
biological systems involve simple entities which lack strategic
complexity. This prevents them to adopt strategies that take into
account the history of their interactions with other entities.
Yet, remarkably, cooperation is present in theses systems, as
supported by evidence
\cite{soares2008cleaning,mehdiabadi2006social,faaborg1995confirmation}.

In the context of decentralized wireless networks, we will be
interested in design rules which are simple enough to be
implemented by communication nodes with limited processing
capabilities, yet powerful enough to promote cooperation and to
yield global energy efficiency. This is in contrast to most of the
present approaches which rely on complex algorithms and reputation
tables in order to enforce cooperation in the network
\cite{zhong2003sprite,buchegger2002performance,michiardi2002core}. 
Our objective is to promote cooperation by relying on
simple strategies, i.e. by imposing a limited set of rules which
mimic the principles of evolution, and let the systems
evolve in time. 

The question we ask is the following: Can cooperation arise in
communication networks by evolution? If yes, which mechanism
should be at work? It seems that network reciprocity is a
promising candidate for promotion of cooperation in communication
networks. Indeed, when wireless networks are described as graphs,
an analogy can be drawn with populations which are not well mixed.
The reason for this is that, given a power constraint, one user
can interact only with the nodes which are in the range of its
transmission, forming a cluster of potential cooperators.

In order to investigate the effects of the application of this
kind of mechanism to wireless communication networks, we define a
relatively simple network model which, however, captures both the
essence of wireless communication networks and the graph models
used in evolutionary game theory. We simulate the emergence of
cooperative behavior in a communication network in order to
explore whether rules such as natural selection can favor
cooperation in these networks.

\subsection{Network model}

We model the network as a graph where the users represent the
nodes and the edges are related to interactions between them.
The objective of each network node is to be power efficient, i.e
to minimize the amount of power it spends for packet transmission.
As in game theory, we assume two types of nodes, cooperators and
defectors. Additionally, we make the following assumptions. First,
we assume that the power decays as a power law function of the
distance to the transmitter (source). Hence, if the transmit power
is $P_T$, the power received at distance $d$ from the transmitter
is
\begin{equation}
\label{eq:power_law}
P_R=\frac{P_T}{Kd^\alpha},
\end{equation}
where $\alpha$ depends on the propagation characteristics of the
area (urban, suburban, rural, etc.) and $K$ is an arbitrary
constant. Typically, $\alpha$ takes values in the range
$2\leq\alpha\leq 4$. Due to the power law decay, the presence of
cooperators might be beneficial with regard to the energy
efficiency. For simplicity we assume either one-hop retransmission
(in the presence of a cooperator), or direct communication (in the case
when there are no cooperators willing to retransmit the packet).

A time division multiple access (TDMA) approach is used where the
nodes take turns in transmitting their packets (no frequency
reuse). We divide the time scale in time slots of equal duration
and assume that one transmitter/receiver pair is activated at
random in each time slot. This multiple access scheme is known
to be
optimal, at least in first approximation, from a minimum energy per bit perspective. Although this assumption simplifies the network analysis, it
may be regarded as restrictive, as it does not include interference/collisions. 
Nevertheless, we expect that this simple scenario will be able
to capture the essence of the cooperative behaviour of the users and that, from the perspective of the investigated phenomenon (emergence of cooperation), the simulation results will be
a reasonable indicator of the network behaviour in the more general case. A precise study of the phenomenon in the scenario which embraces simultaneous transmission is a topic of current work. 
Under these assumptions, we will be essentially interested in the
total power consumed by the network over time.

Let us say that during one time slot user A needs to transmit to user
B and that the power it uses for direct transmission is $P_{D}$.
As a result of the propagation effects, the received power at user B
is $P_{R}=P_D/\left(K d_{AB}^\alpha \right)$. We define the
signal-to-noise ratio at the receiver as $SNR=P_{R}/\sigma^2$,
where $\sigma^2$ is the noise variance. We say that the
transmission is \textit{successful} if the signal-to-noise ratio
at the receiver exceeds a certain threshold, $SNR\geq
SNR_0=P_{R_0}/\sigma^2$, which is required for reliable reception.
In other words, in order to have a successful transmission, the
node A should transmit with power $P_{D}\geq Kd_{AB}^\alpha
P_{R_0}$. In the first instant, for simplicity, we assume perfect
power adaptation (which should be justified in general) and assume
that the node A adjusts the transmit power to the distance
$d_{AB}$, such that it meets the receive SNR requirement exactly.
This is, of course, a simplification, since for this adaptation to
work, A should know the network topology (the distance to B,
$d_{AB}$) or, at least, to have a feedback from B about the
receive SNR such that it can adjust the transmit power $P_{D}$.

\begin{figure}[!htbp]
\begin{center}
\includegraphics[scale=0.25]{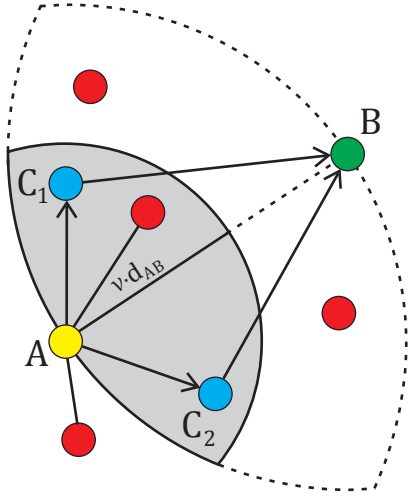}
\caption{Typical communication scenario: yellow circle - transmitter ; green circle - receiver; blue circles - cooperators; red circles - nodes that do not participate in the communication; dashed line - the area where intermediate nodes are located; full line - the area where the cooperators for the transmission/receiver pair A/B are located; }
\label{fig:Intermediate_nodes}
\end{center}
\end{figure}

We say that a node $C$ is in \textit{range of} A, or
\textit{connected} to A, if it can "hear" A's transmission to B.
Under the assumed power adaptation, a node C is in the range of A
if $d_{AC}\leq d_{AB}$. On the other hand, the retransmission only
makes sense if $d_{CB}\leq d_{AB}$. Hence, potential cooperators
will be located in the area described by
\begin{align}
\label{eq:intermediate}
  d_{AC} &< d_{AB};\nonumber\\
  d_{CB} &< d_{AB}.
\end{align}
A node C which fulfills (\ref{eq:intermediate}) is called
\textit{intermediate} node. The equality in
(\ref{eq:intermediate}) has been left out in order to avoid
defining B as an intermediate node. In the time period when node A
transmits data to node B, there may be several intermediate nodes.
In Fig.~\ref{fig:Intermediate_nodes} the intermediate nodes are
located in the area enclosed with a dashed line. 
 In order to decrease the transmission cost for user A,
 we further reduce the region where the possible cooperators are
 found by introducing a parameter $\nu \in (0,1)$. With this convention,
 instead of engaging all cooperators from the area defined
with (\ref{eq:intermediate}) in the retransmissions, we account
only for those in the area
 enclosed by full line. This area includes the nodes at reduced distance $\nu \cdot d_{AB}$
 from the transmitter. 

The relations between the nodes are represented as edges in the
graph. We note that this means that each transmission from A to B
is associated with one directed network graph. Since in real
networks there are simultaneous transmissions between different
pairs of users, at each time slot the networks is actually
described by a \textbf{set} of different directed graphs, rather
then a single graph. Depending on the
users activity, the set of graphs also changes over time. 

Now, if an intermediate node C decides to help A in the
transmission, i.e. to cooperate, it will retransmit the signal
received from A to B.
The benefit that node A obtains from the cooperative act of C is that it can decrease the transmit power to a value lower then the power
required for direct transmission, $P_{I}\leq P_{D}$, where the
subscript $I$ stands for indirect transmission (transmission when
cooperators are involved). In this context, we can define the
benefit of the cooperative act as $b=P_{D}-P_{I}$.

In general for a given transmitter/receiver pair there can be multiple cooperators. In this case they can either share the cost for cooperation or let one cooperator pay the overall cost.
For simplicity, we propose that only
the closest cooperator to B retransmits the signal. Different approaches to cost sharing will be left for future study. Without loss
of generality, let $k$ be the index for which $d_{C_kB}$ is
minimal. As elaborated before, we take that the power received at
B should be exactly the minimal one required for successful
transmission, $P_{R_0}$. In this case the cost that the cooperator
$k$ (the closest one) pays is
\begin{equation}
P_{C_k}=P_D(C_k,B),
\end{equation}
while the other potential cooperators pay no cost.

\begin{figure}[!htbp]
  \centering
  \includegraphics[width = 0.23\textwidth]{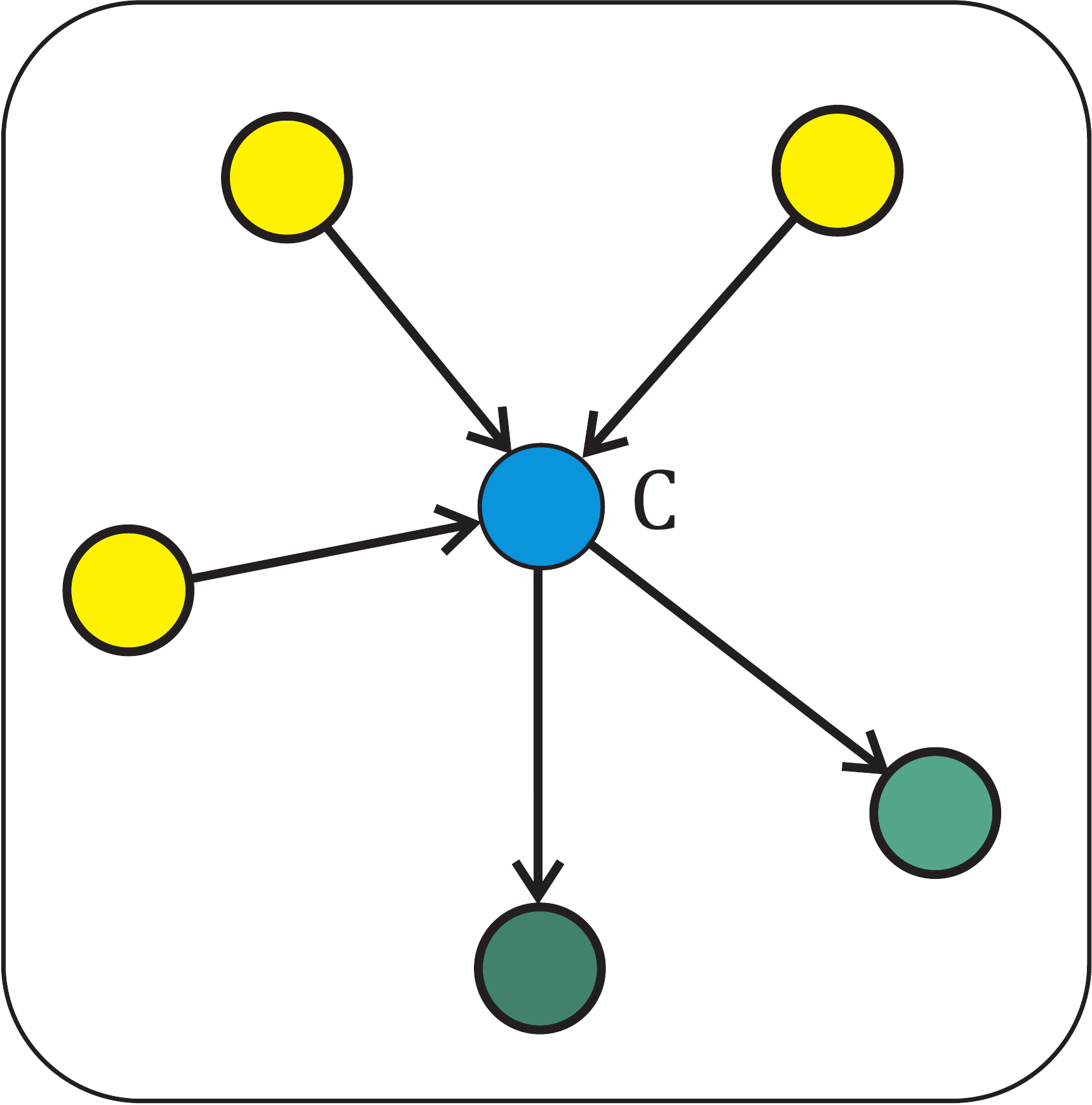}
  \caption{Incoming and outgoing edges for node C.}
  \label{fig:node_c_in_out_edges}
\end{figure}

Let us now change the perspective and look at the network at time
slot $n$ from the viewpoint of a single node (user), node C for example. For the node C we distinguish between
\textit{incoming} and \textit{outgoing} edges as depicted in Fig.~\ref{fig:node_c_in_out_edges}. The outgoing edges
are associated with the nodes which are in the range of C, when C
transmits its own packets.
We take that the number of incoming edges to node $C$ is $L$, out of
which $J\leq L$ are active (associated with ongoing
transmissions).
According to this model, the total power that C
spends for cooperation, $P_C$, is a function of $J$, $P_C=P_C(J)$.
On the other hand, the power that C spends for its own
transmission is either $P_I$ or $P_D$.

If the node C is in the range of a transmitting node, let
us say A for example, then there is an incoming edge to C,
outgoing from A. We note that by adapting this model, we allow
that at one time slot the node C can be in the range of several
other nodes and also help several of them in the transmission. In
reality this can be done by performing a kind of multiplexing at
the nodes, for example by using spread-spectrum sequences to
distinguish between the different users.

Following the analogy with biology, we will define fitness of the
individual nodes. Intuitively, the fitness has to be related with the energy consumption of the
individual nodes. Ideally, the appropriate fitness function has to
be simple enough to be evaluated locally (possibly without
requiring complex processing and memory), but also rich enough to capture the essence of wireless
transmission and network dynamics. In addition, we recall
that the strategies of the individual nodes (in game-theoretic
sense) are adopted with respect to the fitness function. In this
sense, the choice of the fitness function is related to
the game-theoretic analysis of the different strategies.

We will define two discrete time scales,
according to which the fitness will be evaluated. 
The fitness function is evaluated at the end of a block of
duration $T$ slots. The network performance is observed over $N$
such blocks (iterations).

Since we have two time scales, we introduce two indices, $t$ and $n$,
where $t\in\left\{0,1,\ldots,T\right\}$ indicates the time slot
and $n\in\left\{0,\ldots,N\right\}$ indicates the iteration. The
fitness is then a function of $n$ and $t$, $F=F(n,t)$. We note
that, in order to be consistent with the definition of $t$, we
denote the initial iteration (of duration $T$) as the $0$-th
iteration. Additionally, we denote the initial fitness as
$F(0,0)=F_0$.

Now, let us define
\begin{equation}
\Delta f(n,t)=F(n,t)-F(n,t-1),\:\: t=1,\ldots,T ,
\end{equation}
which measures the difference in the fitness evaluated at two
consecutive time instants $t-1$ and $t$, of the $n$-th iteration.
In our model, $\Delta f(n,t)$ for the network node C is
defined as
\begin{equation}
\Delta f(n,t)=-\alpha \left( 1 - \beta \right) \left[ P_D-P_I \right] - \gamma \delta P_C(J)
\end{equation}
where $\alpha, \beta, \gamma, \delta \in{\lbrace0,1\rbrace}$ are parameters
which indicate packet transmission and the presence of cooperators and
defectors. In particular, $\alpha=1$ when C has a packet to
transmit; $\beta=1$ when C has at least one cooperator as a
neighbor; $\gamma=1$ when C is connected to at least one active
node at that time instant; and $\delta=1$ corresponds to C being a
cooperator (otherwise the parameter values are zeros). We note that the above parameters are also functions
of $n$ and $t$. However, whenever there is no ambiguity, and in
order to simplify the notation, we will skip these indices.

Having introduced $\Delta f(n,t)$, we can define the fitness of
the node C in the following way
\begin{align}
\label{eq:fitness}
F(0,0)&=F_0,\nonumber\\
F(n,t)&=F(n,t-1)+\Delta f(n,t),\nonumber\\
F(n+1,0)&=F(n,T)
\end{align}
where $n=0,1,\ldots, N-1$ and $t=0,1,\ldots, T$. In addition, we define the quantity
\begin{equation}
\Delta F(n)=F(n,T)-F(n-1,T),\:\: n=1,\ldots,N
\label{eq:delta_fitness}
\end{equation}
to be the change of fitness between two consecutive iterations,
$n-1$ and $n$. As we will see, this quantity will be relevant for
the definition of the different strategies.






\subsection{A Game-theoretic Approach}
 Game-theoretic approaches to modeling of phenomena assumes the existence of some quantity -- utility, or benefit -- that
 units in the system try to maximize. In some scenarios the agents may choose to help the others, i.e. to cooperate -- this is modeled
 by the cost they pay for the cooperation. Some agents choose their strategy to be selfish, i.e. they defect, and thus avoid any costs.
 The cost of cooperative act implies that the cooperators will have smaller fitness than the defectors.
 Thus natural selection of the fittest favors defectors. However, there are observations and theoretic
 analyzes of cases when cooperation persists -- there is at least a fraction of cooperators present in the population.

In this work we study four different strategies of cooperative behavior of nodes in communication networks.
According to our approach we assume that all network nodes adopt
the same strategy during the simulations. This approach certainly
does not cover some more general scenarios, for example one
where the individual nodes are able to choose their strategy at
random, or according to some rule. Nevertheless, we expect that
the results from our analysis will fairly well indicate the
general trend and, as such, will be useful in the evaluation of
the fundamental limits of the energy efficiency in decentralized
networks.

The first strategy addresses the trivial case when there is no
cooperation between the nodes, i.~e. all nodes are defectors. We
denote this strategy by DEF. The second strategy corresponds to
the case where all nodes cooperate and will be denoted as
COOP. It corresponds to a centralized scenario where
the cooperation is in a way enforced in the network.
The main effect we expect from cooperation among the nodes is a
decrease in the total energy consumption. The simulation results show that when all nodes cooperate the total energy
consumption is reduced by as much as 60\% as opposed to the case
when all nodes defect.

As already discussed, for us of greater interest are
strategies which are appropriate for the decentralized scenario at
hand, where the individual nodes decide whether to cooperate or
not based on their individual fitness. In the decentralized
scenario that we propose, at the end of each iteration the network
nodes decide whether to cooperate or defect in the next iteration.
As we do not assume nodes with strategic complexity, we will
concentrate on simple strategies. According to this approach, at
the end of the $n$-th iteration each node makes the decision only
based on the change in the fitness $\Delta F(n)=F(n,T)-F(n-1,T)$,
as defined in (\ref{eq:delta_fitness}).

We will distinguish between two simple and intuitive strategies
for the decentralized scenario. According to the first one, if the
node observes an increase in the fitness, $\Delta F(n)>0$, it will
retain the previous status in the next iteration. If, on the other
hand, the node observes a decrease in the fitness $\Delta F(n)<0$,
the node will change its behavior, i.~e. a cooperator will become
a defector and vice versa. We observe that from the perspective of
a single node, the game resembles the repeated prisoner's dilemma
\cite{axelrod1981evolution}. In this regard, the above described strategy
corresponds to the well known \emph{win-stay, lose-shift} and is
based on the simple idea of retaining the previous status when the
node is doing well, but changing it otherwise. In the remaining of
the text we refer to this strategy as WSLS.

According to the other strategy for the decentralized scenario,
the node will decide to cooperate in the next iteration if it
observes an increase in the fitness, $\Delta F(n)>0$. Otherwise,
it will defect. According to this strategy, a defector will become
cooperator and a cooperator will stay cooperator, if $\Delta
F(n)>0$. Otherwise, the node will choose to defect. We note that
the increase in fitness reflects the average behavior of the
adjacent nodes, in the sense that the reason for the fitness
increase is the cooperative behavior of some of the adjacent
nodes. In the context of the repeated prisoner's dilemma, this
strategy resembles the \emph{tit-for-tat} strategy which is based
on the idea of mimicking the other node(s) behavior in the
previous turn. This means that the node will become cooperator
only if it observes cooperative behavior of other nodes which is
reflected in the increase in the fitness. In the remaining of the
text, we will refer to this strategy as TFT.

\section{Description of the Experiment}
\label{sec:experiment_setup}

\subsection{Simulation setup}
Having defined the different strategies, the aim of the
simulations will be to evaluate and compare the performance of all
strategies in terms of both the energy consumption of the
individual nodes and the global energy consumption in the network.
As already discussed, although we would like to see the global
energy consumption decreased, we would also like to have the
energy consumption distributed as uniformly as possible between
the different nodes.

Our simulation is set up as follows. We place $M$ wireless nodes
at random in a circle of radius $r$, according to a uniform
distribution. Since we are interested only in the relative
performance
of the different strategies and not in the absolute value of the consumed energy,
we normalize the circle radius to a single distance unit, $r=1$. 
The nodes send their messages during time slots of fixed duration
(same for all nodes), where in each time slot exactly one
transmitter/receiver pair is activated at random. We group the
time slots in blocks of length $T$, where we denote the block of
$T$ slots as one iteration. The network behavior is observed over
$N$ iterations. 


We consider two simulation scenarios. In the first scenario we
assume that either all nodes are defectors (DEF strategy), or
cooperators (COOP strategy). In the second simulation scenario we
start by assuming that in the initial iteration all users are
defectors. At the end of the initial iteration we choose one user
at random and change its behavior from defector to cooperator. At
the end of each next iteration the nodes perform a
fitness update and perform a comparison with the fitness at the
end of the previous iteration, as given by (\ref{eq:delta_fitness}).
According to the change of the fitness defined in
(\ref{eq:delta_fitness}), the users determine their behavior during the
next iteration (cooperate or defect) according to the TFT or the
WSLS strategy. Once again, we note that in our setup all users
follow the same strategy during all iterations.

In both scenarios we are interested in the individual and in the
total energy consumption during the $N$ iterations. Since we are
interested in the relative performance of the different
strategies, for reference we will take the average consumed energy
(power) per node in the case when the nodes adopt the DEF
strategy. Without loss of generality, we will consider this energy
to have value $1$ and express the energy consumption of the
different strategies with respect to this value.

The simulations were performed for the extreme case of the
propagation parameter $\alpha = 4$, for which the cooperation is
most beneficial in terms of energy consumption. To mimic the
simultaneous transmissions present in reality, we have taken a
frame length of $T = 1000$ consecutive time slots that correspond
to 1000 packets. Additionally, the simulations last $N = 1000$
iterations. At last, we have considered $N_t=1000$ different
network topologies obtained by random placement of the nodes in
the unit circle with uniform distribution. The large number of
repetitions has the effect of smoothing the graphical results. The
choice of the numerical values of the simulation parameters are
summarized in Table~\ref{tab:param_vals}.

\begin{table}[!htb]
\caption{Simulation parameters}
\begin{center}
\begin{tabular}{ l | r || l | r }
\label{tab:param_vals}
  Parameter & Value & Parameter & Value \\
  \hline
   $T$ & 1 000 &  $r$ & 1 \\
   $\alpha$ & 4 &  $N$ & 1 000  \\
   $M$ & 30 &  $N_t$ & 1 000 \\
\end{tabular}
\end{center}
\end{table}

\subsection{Scenario 1: DEF and COOP strategy}

When the users adopt the DEF strategy, a transmitter A
communicates with a receiver B by direct transmission. Hence the
cost associated with the transmission is $P_{D}= Kd_{AB}^\alpha
P_{R_0}$, where $P_{R_0}$ is the minimal receive power necessary
for successful transmission, as defined in
Section~\ref{sec:network_model}.

In the case when the users adopt the COOP strategy, for a
transmitter/receiver pair A/B, node A receives a benefit from the
cooperators located in the area defined by
(\ref{eq:intermediate}). 

The benefit is reflected in the fact that A can adopt (reduce) the
transmit power to a value $P_{I}= K \left( \nu \cdot d_{AB}
\right)^\alpha
P_{R_0}$, 
where $0<\nu<1$ is a parameter which determines the reduced range
of A's transmission. 
Once fixed, the parameter $\nu$ is kept constant for all nodes and
during all transmissions. As we will see, the results from the
simulations indicate that the choice of this parameter is
particularly important for the energy consumption in the network.

\begin{figure}[!htbp]
\begin{center}
\includegraphics[scale=0.41]{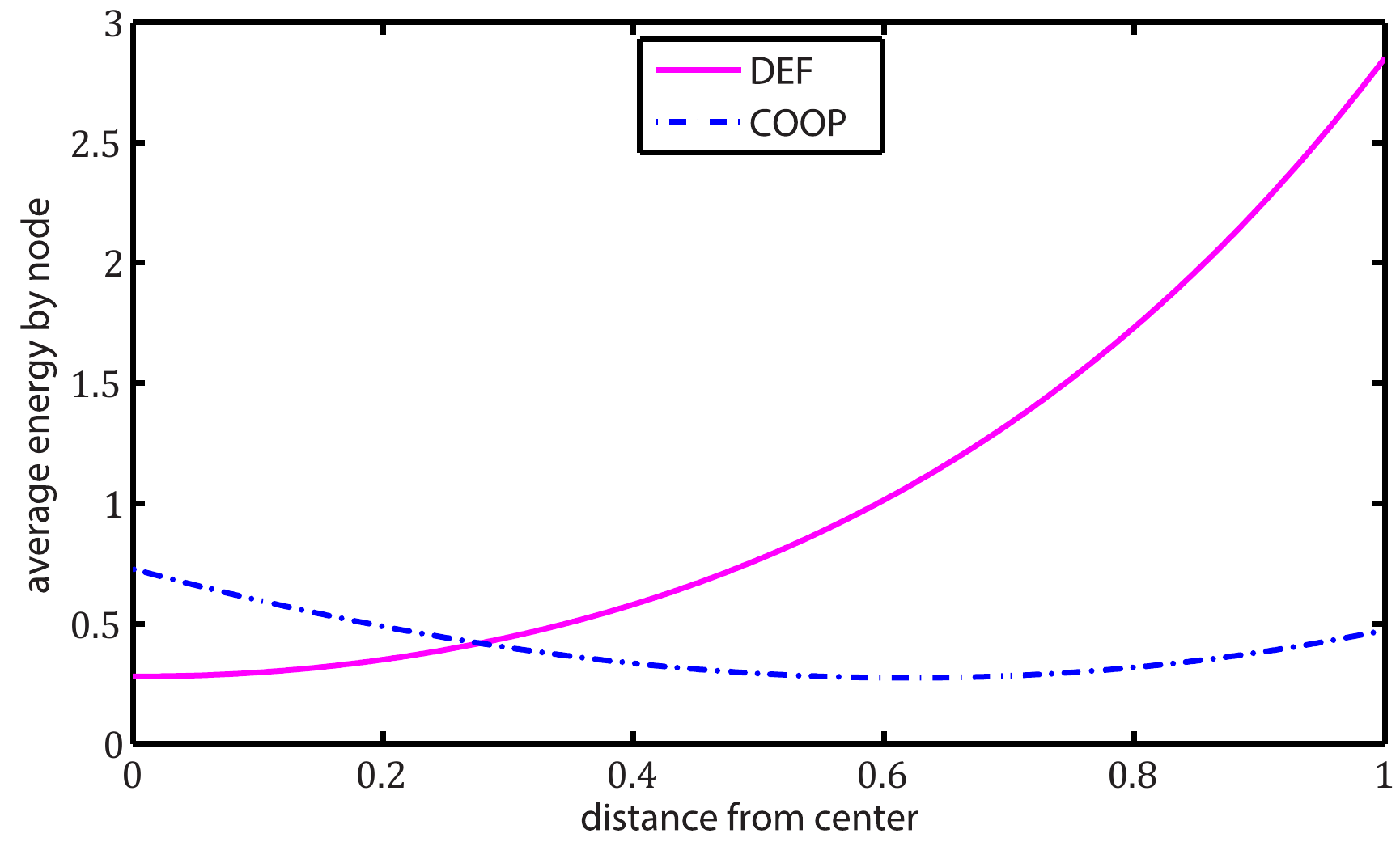}
\end{center}
\caption{Average energy consumption for nodes at different
distances from the center.}
\label{fig:energy_spent_vs_r_def_coop}       
\end{figure}

The main effect we expect from cooperation among the nodes is a
decrease in the average energy consumption. The results from the
simulations show that when all nodes cooperate the average energy
consumption is reduced by as much as 60\% as opposed to the case
when all nodes defect, as indicated in Table~\ref{tab:mean_std}. Apart from the
reduction in average energy consumption, cooperation leads to a
more-fair energy consumption among the individual nodes. Indeed,
when there is no cooperation the nodes which are located further
from the center are at a disadvantage as the average distance to
the rest of the nodes is larger compared to the nodes which are
located near the center, leading to increased energy consumption
for transmission. The introduction of cooperation lessens this
imbalance to some extent. An illustration of this effect is shown
in Fig.~\ref{fig:energy_spent_vs_r_def_coop}. As we can see, the
introduction of cooperation balances the amount of energy spent by
the individual nodes, and decreases the effect of the network
topology on the individual energy consumption.

\subsection{Scenario 2: TFT and WSLS Strategy}
As discussed, in this simulation scenario we start by assuming
that in the initial iteration all users are defectors. At the end
of the initial iteration we choose one user at random and change
its behavior from defector to cooperator.

For one transmitter/receiver pair A/B, the nodes which fulfill
(\ref{eq:intermediate}) are potential cooperators. 
If there is a cooperator present, it retransmits the signal to
user B. Now, based on the estimation of the receive SNR, the
receiver B informs A via feedback (which can also be quantized)
that there are cooperators present. With this in mind, the sender
A
adopts (reduces) the transmit power to a value
\begin{equation}
P_I = K\left(\nu \cdot d_{AB} \right) ^\alpha P_{R_0},
\end{equation}
We note that by introducing this heuristics, the sender A does not
have to know the identity of the cooperators in each iteration. We
recall that after a period of $T$ time slots (one iteration), the
nodes calculate their fitness and compare it with the fitness at
the end of the previous iteration, as given by
(\ref{eq:delta_fitness}).


The simulation results indicate that the choice of the parameter
$\nu$ is crucial for the energy consumption. We fix this parameter to $\nu=0.39$, a value for which the total energy consumption is approximately minimal, for all strategies, as indicated by the performed simulations.  

\subsection{The effect of cooperation}

\begin{figure}[!htbp]
\begin{center}
\includegraphics[scale=0.42]{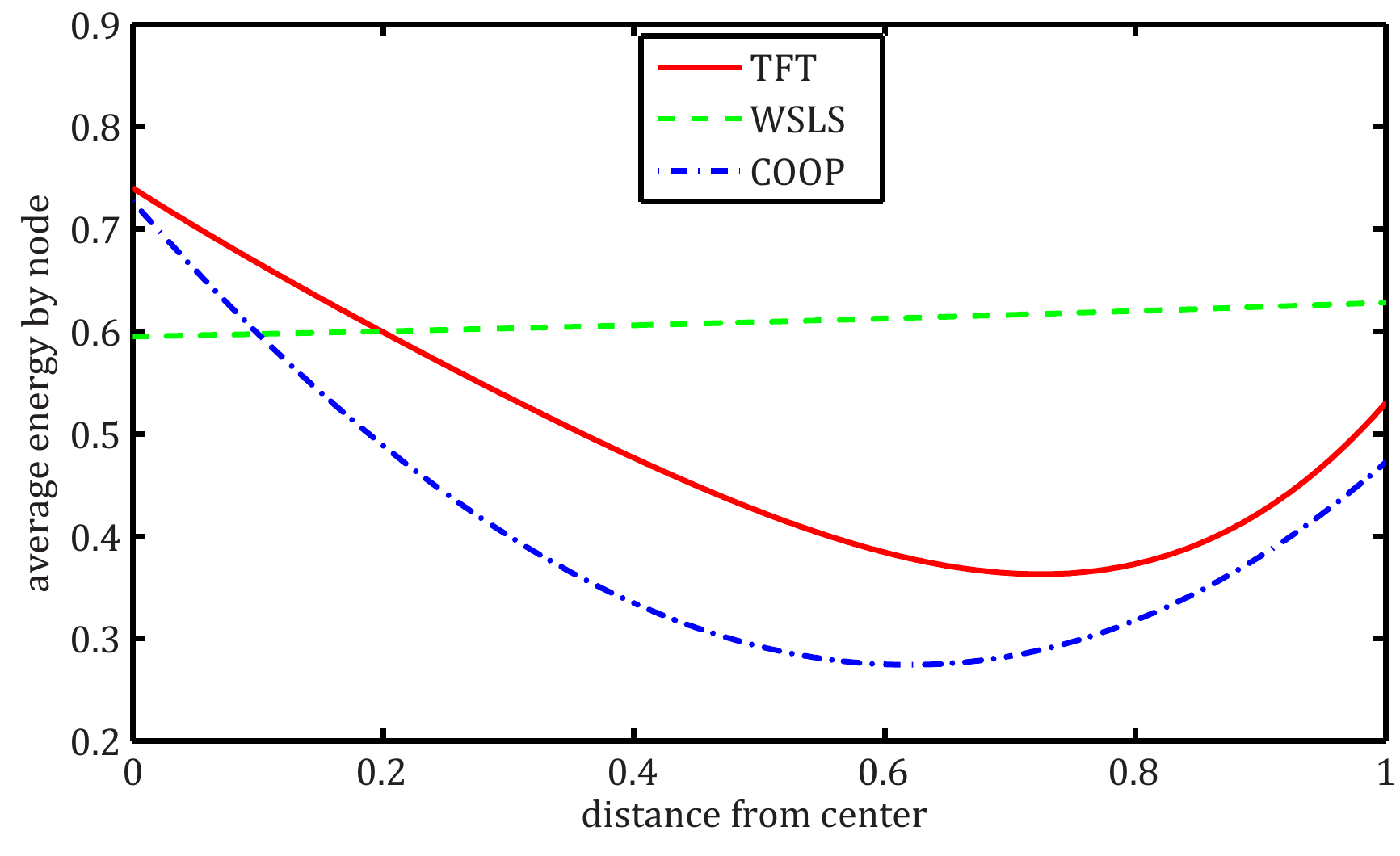}
\end{center}
\caption{Average energy consumption for nodes at different distances from the center.}
\label{fig:energy_spent_vs_r_wout_def}       
\end{figure}

Although the simulation results (as expected) show that the COOP
strategy yields a minimal total energy consumption among all four
strategies, this is not the optimal strategy from the perspective
of all of the individual nodes. Indeed, as
Fig.~\ref{fig:energy_spent_vs_r_wout_def} indicates, the
simulation results show that the adoption of the WSLS strategy
yields lower energy consumption for the nodes closer to the
center. Further the WSLS strategy yields the most balanced energy
consumption as function of the geographical distribution of the
nodes.
The difference in the balance of energy consumption can be inferred from the standard deviation values given in Table~\ref{tab:mean_std}.

\begin{table}[!htbp]
\caption{Mean and standard deviation of the average energy consumption $E$}
\label{tab:mean_std}
\begin{center}
\begin{tabular}{ l | r | r }
  strategy & mean($E$) & std($E$) \\
  \hline
   DEF  & 1.00000 & 0.72150 \\
   COOP & 0.39755 & 0.13093 \\
   TFT  & 0.48858 & 0.11559 \\
   WSLS & 0.60966 & 0.01671 \\
\end{tabular}
\end{center}
\end{table}

Let us recall that the population starts with only a single
cooperator while all other nodes are defectors. The simulations
show that with time cooperation spreads through the network. As a
single node changes its behavior frequently during a single
simulation (depending on the fitness evaluation and on the
strategy), we calculate the frequency of cooperation as 
\begin{equation}
\label{eq:freq}
f(i) = \frac{ \mbox{number of iterations when node }  i \mbox{ is cooperator}  }{ \mbox{total number of iterations} }
\end{equation}

\begin{figure}[!tbp]
\begin{center}
\includegraphics[scale=0.42]{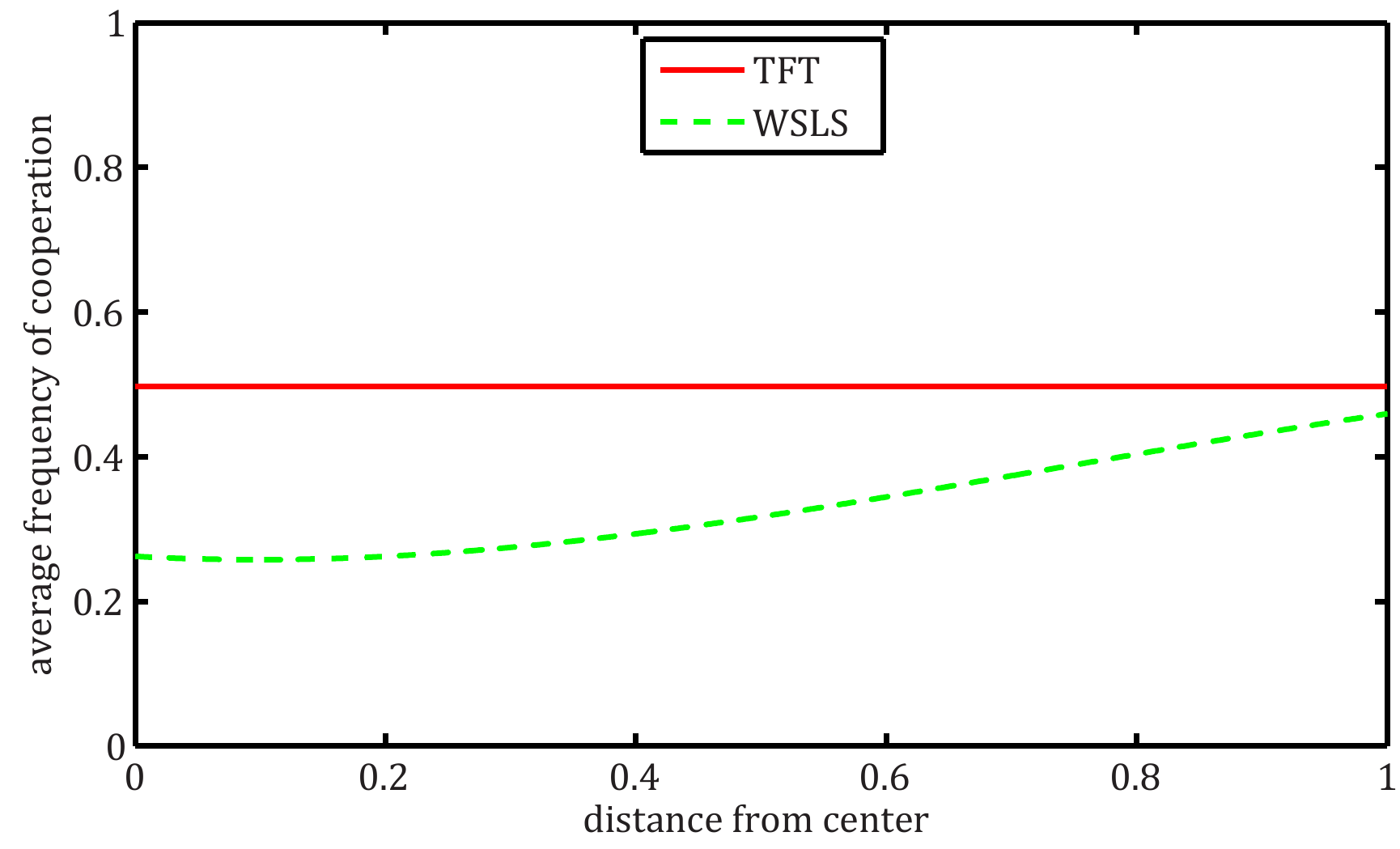}
\end{center}
\caption{Average frequency of cooperation for nodes at different distances from the center.}
\label{fig:percent_coop_vs_r}       
\end{figure}

The results of the simulations show that the frequency of
cooperators is fairly constant (around 50\%) for the different
placement of the nodes for the TFT strategy, as depicted in
Fig.~\ref{fig:percent_coop_vs_r}. The explanation for this effect
is that when the nodes use the TFT strategy, they are more prone
to change their behavior from defectors to cooperators and vice
versa. We recall that the frequency of cooperators is averaged
over 1000 different network topologies, and therefore the constant
frequency of cooperators does not come as a surprise. As for the WSLS strategy, it is intuitive to expect that nodes
closer to the center will be less incentive to cooperation,
compared to those which are further away. 

Fig.~\ref{fig:percent_coop_vs_r} can help us understand the balanced energy consumption present in the WSLS scenario. A key observation is that nodes which are positioned closer to the center are cooperators less frequently then others. This reduces their individual energy consumption. Additionally, the absence of cooperators close to the center shifts the cooperation burden to the outer regions. The peripheral nodes thus experience an increase in their energy consumption. Overall, these two effects combined reduce the differences in the individual energy consumption witnessed in TFT and produce the flat curve shown in Fig.~\ref{fig:energy_spent_vs_r_wout_def}.

\section{Conclusions}
\label{sec:conclusion} We investigated the mechanisms for
promotion of cooperation in decentralized wireless networks. The
approach was motivated by recent results in evolutionary biology
which suggest that cooperation can be favored by natural
selection, if certain mechanisms are at work. We modeled the
wireless network as a graph, where benefits and costs were
associated with the strategy that the network users follow. In
game-theoretic spirit, the nodes based their behavior on
calculations of their energy spending.

We presented numerical study of cooperative communication
scenarios based on simple local rules, which is in contrast to
most of the approaches in the literature which enforce cooperation
by using complex algorithms and require strategic complexity of
the network nodes. We considered four strategies for the users'
behavior and observed the energy consumption of the individual
nodes and the network in total. In two of the strategies, TFT and
WSLS, we considered that the nodes determine their behavior only
based on individual fitness. The simulations showed that even selfish decision making 
of the nodes can lead to emergence of cooperation. These
observations served as indicator that uncomplicated local rules,
followed by simple fitness evaluation, can generate network
behavior which yields global energy efficiency.

We recall that here we adopted the convention that the same
strategy was used by all users in all iterations. In a future
version of the work, we will consider the case where each of the
individual users is allowed to choose its own strategy at every
iteration. As discussed, the results from the simulations indicate
that, depending on the node distance from the center, distinct
nodes could find optimal to follow different strategies. It is
expected that this analysis will bring valuable insights in the
dependencies between the choice of optimal strategy for the
individual users and the network topology.

In addition, it will be interesting to evaluate the network
behavior in the case when the nodes have buffers with limited
energy capacity, under a particular random arrival process. This
is in contrast to the here addressed scenario where we assumed
that the nodes have infinite-length buffers. We expect that the
adoption of this more realistic assumption will influence both the
behavior of the individual nodes and the way energy is consumed in
the network. This more general approach also includes the energy
harvesting scenario where the nodes harvest energy quants from the
environment according to some arrival process.

\section*{Acknowledgment}
The authors acknowledge the support of the German Research
Foundation (Deutsche Forschungsgemeinschaft - DFG), via the
project Li 659/13-1.

\bibliographystyle{IEEEtran}
\bibliography{trudovi}

\end{document}